\documentstyle[prb,aps,epsf,twocolumn]{revtex}
\tighten
\begin{document}
\draft 
 
\title{Semiclassical theory of magnetotransport through a chaotic quantum well}
\author{E.~E.~Narimanov$^1$, A.~Douglas ~Stone$^1$, and 
G. S. Boebinger$^2$}   
\address{$^1$Applied  Physics,  
Yale  University,  P.O.  Box  208284,  New Haven CT 06520-8284 \\
$^2$Bell Laboratories, Lucent Technologies, 700 Mountain Ave, Murray Hill,
NJ 07974}  
\date{\today}  
\maketitle  

\begin{abstract}  
We develop a quantitative semiclassical formula for the resonant tunneling
current through a quantum well in a tilted magnetic field.
It is shown that the current depends only on periodic orbits within 
the quantum well. The theory explains the puzzling 
evolution of the tunneling spectra near a tilt angle of $30^{\circ}$ 
as arising from an exchange bifurcation of the relevant 
periodic orbits.
\end{abstract}

\narrowtext
The resonant tunneling diode (RTD) in a magnetic field tilted with
respect to the tunneling direction, has been extensively studied
in recent years as a simple experimental system which manifests
the quantum signatures of classical chaos 
\cite{fromhold,muller,ss,monteiro,ns_long,scars,ns_scars}.
The measured I-V characteristics show resonance peaks which evolve in
a complex manner as magnetic field, $B$, and tilt angle, $\theta$, are
varied\cite{fromhold,muller}.  
The existence and periodicity of these peaks in various
parameter intervals has been associated with the existence of certain
periodic orbits \cite{fromhold} and their bifurcations\cite{ss,muller}.  
The link to quantum mechanics
has been made by intuitive appeals to Gutzwiller oscillations of the
density of states\cite{fromhold}, 
scaling analyses of the exact quantum spectrum \cite{monteiro}, and
the numerical discovery of sequences of wavefunctions scarred by
periodic orbits \cite{scars}.  However, previous to this work, it has
not been shown that periodic orbits indeed determine the quantum tunneling
oscillations in the semiclassical limit.  Indeed, the analogy between 
the tilted well and another
well-studied chaotic system, the hydrogen atom in magnetic field, has
frequently been emphasized \cite{fromhold,monteiro}; however in diamagnetic
hydrogen it has been shown \cite{delos} 
that {\it all} orbits closed at the nucleus
determine the absorption spectrum, {\it not just the periodic orbits}.
Moreover, even assuming the importance of periodic orbits, 
the periodic orbit theory
of the system is sufficiently complex that the contribution of specific
periodic orbits and their bifurcations to the experimental I-V data
has been controversial \cite{fromcomment}. Below we derive
a quantitative semiclassical formula for the tunneling current which
demonstrates that the current is dominated by periodic orbits and apply 
the formula
to previously unpublished data which reveals an interesting {\it exchange
bifurcation} involving four period-two orbits.

In an RTD under a bias voltage $V$, tunneling current 
flows from the emitter state through the double-barriers confining the 
quantum well.  The data presented are from an RTD with a $120$nm 
wide well and experimental details are given in Ref. 2.
When a large magnetic field ($> 1 {\rm T}$) 
is applied, the emitter state is quantized into the
first few Landau levels. The electric field is
normal to the barriers (${\bf E} = E{\bf {\hat z}}$), while the magnetic
field is tilted in the $y-z$ plane, 
${\bf B} = \cos \theta {\bf {\hat z}} + \sin \theta {\bf {\hat y}}$. 

After tunneling into the well through the 
emitter barrier, an electron gains large kinetic energy before
colliding with the collector barrier.   Typically, the electron will
traverse the well hundreds of times before tunneling out
and will lose much of its energy due to optic phonon emission.
Therefore, the tunneling is sequential and the resonances are substantially
broadened by $\hbar/\tau_{opt}$, where the phonon emission time
is $\sim 0.1$ ps \cite{fromhold}.  For describing this limit the
Bardeen tunneling hamiltonian formalism is appropriate \cite{bardeen,payne},
which greatly simplifies the problem.  Using this approach and taking 
into account that the relative barrier widths
are chosen so that charge accumulation in the well is negligible
\cite{muller}, one finds that the current is simply proportional
\cite{explain_emitter_occupation} to the
tunneling rate through the emitter barrier, $W_{e \rightarrow w}$  :
$j = n_e e W_{e \rightarrow w}$,
where $n_e$ is the surface concentration in the emitter layer.

$W_{e \rightarrow w}$ can be calculated from the Fermi Golden Rule with
the coupling matrix element \cite{bardeen,payne} between the wavefunctions
 $\Psi_e$ and $\Psi_w$, corresponding to the
{\it isolated} emitter and {\it isolated} well respectively. In the limit when
the height of the emitter barrier $U_0$ is much larger than the injection
energy $\varepsilon_i$, the cyclotron energy $\hbar \omega_c$ and the 
voltage drop across the barrier, the coupling matrix element 
can be simplified to :
\begin{eqnarray}
M^{e \rightarrow w}_{nk} = \frac{\hbar^2}{m^*} 
\int d {\bf S} 
\Psi^{e}_n(x,y,0) \left. 
\frac{\partial\Psi^{w}_k(x,y,z)^*}{\partial z}\right|_{z=0}
\label{me_approx}
\end{eqnarray}
where the integration is performed over the inner surface of the 
emitter barrier $z = 0$.

Due to the translational invariance in the x-direction, the classical
dynamics within the well can be reduced to two degrees of freedom,
$y,z$, \cite{muller,ss,monteiro,ns_long} with an effective potential
$V\left(y,z\right)$. The tunneling
rate, which is proportional to the square of the matrix element
(\ref{me_approx}),
 can be expressed in terms of the related Green function 
$G(y_1,z_1=0;y_2,z_2=0;\varepsilon)$. 
This Green function is then replaced by its
semiclassical approximation \cite{gutzwiller}, which is determined
by {\it all} classical trajectories connecting the points $(y_1,0);(y_2,0)$.
Defining $y=(y_1 + y_2)/2, \Delta y = y_1-y_2$, it is convenient to
introduce the Wigner transform of the emitter wavefunction, 
$f_W\left(y, p_y\right)= h^{-1} \int d \Delta y 
\Psi_e\left(y - \Delta y/2, 0 \right)
\Psi_e^*\left(y + \Delta y/2, 0 \right) 
\exp\left( i p_y \Delta y/ \hbar \right)$. 
Since the emitter state $\Psi_e$ involves only the few lowest single-particle
levels, it can be
calculated accurately using a variational approach.

We then obtain for the oscillatory part of $W_{e \to w}$:
\begin{eqnarray}
W_{\rm osc} & = &
\int dp_y \int dy \ f_W\left(y,p_y\right)
\sum_\gamma 
\frac{\left(p^\gamma_z\right)_i \left(p^\gamma_z\right)_f}
{ \left(m^*\right)^2 }
\int d \Delta y
\nonumber \\
& \times & 
\Re   \left\{
\frac
{8 D_\gamma^{1/2}}
{\sqrt{2 \pi \hbar i}  }
\exp\left[ 
- \frac{t_\gamma}{\tau_{\rm opt}}
+
i
\frac{
S_\gamma
- p_y \Delta y}
{\hbar}
\right]
\right\}
\label{unclosed}
\end{eqnarray}
where $\gamma$ is the index of the classical trajectories, 
$S_\gamma \equiv 
S_\gamma\left(y - \Delta y, 0; y + \Delta y,0;\varepsilon\right) $ 
is the action
integral,  $ t_\gamma$ is the traversal time and 
$D_\gamma$ is the (complex) amplitude \cite{gutzwiller}.
The momenta $p^\gamma_i$ and $p^\gamma_f$ correspond to 
respectively the initial and final points of the trajectory  
$\gamma$.
The factor $\exp\left( - {t_\gamma}/{\tau_{\rm opt}} \right)$
represents the effect of phonon emission and suppresses the
contributions of trajectories longer than $\tau_{opt}$ (corresponding
to $\sim 4$ bounces with the collector).

The next step depends on the properties of the emitter wavefunction.
If this wavefunction were well-localized spatially on the scale
of the electron wavelength within the well, then effectively it
would provide a delta-function source and the resulting formula
for the tunneling rate would involve all closed orbits starting
exactly at the injection point, as in the atomic case \cite{delos}.
However, as noted above, the emitter state is a linear combination of
the first few Landau levels and its Wigner function
$f_W(y,p_y)$ has a typical spatial spread of order
the effective magnetic length, $l_B \equiv \sqrt{\hbar/eB\cos\theta}$,
centered at the ``injection point'' $y = y_i$, and falls off rapidly
outside this interval.  Hence
$\Delta y \sim \hbar/p_y \sim l_B \sim \sqrt{\hbar}$, and 
in the semiclassical limit $\hbar \to 0$ 
one can expand $S_\gamma/\hbar$ to second order 
in powers of $\Delta y$ near the action of the closed orbit.
The function $S_{\gamma}(y,y)/\hbar$ is 
rapidly varying on this spatial scale as $\hbar \to 0$ and its
stationary points correspond to the {\it periodic} orbits 
\cite{gutzwiller}.
Let $y_{\mu}$ be a point of contact with the emitter 
for a given {\it isolated } periodic orbit
(labelled by the index $\mu$), then 
all nearby closed orbits
can be represented to the correct order by quadratic expansion
in $\delta y = y - y_{\mu}$.
If $y_{\mu}$ is within $l_B $ ($\sim \hbar^{1/2}$) of the injection point
$y_i$, then this orbit will contribute substantially. In this
case the quadratic expansion correctly approximates the value of
$S_\gamma\left(y,y\right)$. If $|y_\mu - y_i| \gg l_B$
then this expansion is inaccurate near $y_i$, 
but the contribution of such orbits is
negligible due to rapid variation of the phase 
$S_\gamma\left(y,y\right)$. One then finds
\begin{eqnarray}
W_{\rm osc} & = & 
\frac{16}{m^*} \sum_{\mu}  
\frac{p^{\mu}_z \exp\left( - \frac{T_\mu}{\tau_{\rm opt}} \right) }
{\sqrt{|m_{11}^{\mu} + m_{22}^{\mu} + 2|}}
\int dy \int dp_y  
\nonumber \\
&\times&  
f_{W}^{(e)}\left(y,p_y\right)
\cos\left[ 
\frac{S_\mu}{\hbar} - \frac{\pi n_\mu}{2}
+  Q_{\mu}\left( \delta y,\delta p_y\right)
\right] 
\label{w_osc}
\end{eqnarray}
where
\begin{eqnarray}
Q_{\mu} & = & 
\frac{2}{\hbar}
\frac
{
m_{21}^{\mu} \left(\delta y\right)^2 
+
\left(m_{22}^{\mu} - m_{11}^{\mu}\right)
\delta y \delta p_y 
-
m_{12}^{\mu}
\left(\delta p_y\right)^2
}
{
\left( m_{11}^{\mu} + m_{22}^{\mu} + 2 \right) 
}, \nonumber
\end{eqnarray}
$\delta p_y = p_y - \left( p_\mu \right)_y$,
the integer $n_\mu$ is 
the topological index \cite{gutzwiller} of the periodic orbit,
and the 
$2\times2$ monodromy matrix\cite{gutzwiller} $M = (m_{ij})$  is 
calculated at the emitter barrier.  Therefore we have shown that
the tunneling current depends only on the periodic orbits.

The summation in (\ref{w_osc}) is performed over all isolated 
periodic orbits, both stable and unstable.
Near a stable orbit, the classical motion is regular and the trajectories
are confined to invariant tori in the phase space.  One can still obtain
discrete quantized energy levels from such regular motion by insisting
that action integrals over the torus be quantized.  This leads to a
sequence of eigenfunctions localized on the tori of the stable islands
\cite{miller,gutzwiller}. In contrast, the classical
motion near an unstable orbit is chaotic with no locally conserved 
actions.  As is now well-known, such motion cannot be semiclassically
quantized to yield discrete energy levels.  Therefore 
one would expect a qualitative 
difference between the contributions of stable and unstable orbits
in Eq. (\ref{w_osc}). This difference,
can be displayed explicitly by performing the
summation over repetitions of the primitive periodic orbits. This summation 
can be performed exactly, yielding:
\begin{eqnarray}
W & = & \frac{8}{m^*} \sum_\mu \left(p_\mu\right)_z \sum_\ell
\Delta\left(\frac{T_\mu}{\tau_{\rm eff}^\mu},
\frac
{
S_\mu\left(\varepsilon_\ell\right)
}
{\hbar}
- \frac{\pi n_\mu}{2}
\right)
\nonumber \\ 
& \times & 
\int dy \int dp_y f_W^e \left(y, p_y\right)
g_\ell^{\mu,\pm} \left(y, p_y\right)
\label{w_po}
\end{eqnarray}
where
$\Delta\left(\sigma,\rho\right) = 
\frac{\sinh\left(\sigma\right)}
{\cosh\left(\sigma\right) - \cos\left(\rho\right)}$, 
and the index $+$ or $-$ denotes stable or unstable orbits.
The quantity $\hbar/\tau_{\rm eff}^\mu$ is an effective level-broadening
which differs in the two cases.

(i) {\it Stable orbits} Here we expect the level-broadening
to arise only due to phonon emission and indeed we find
$\tau_{\rm eff}^\mu = \tau_{\rm opt}$.  If phonon scattering is
absent, $\tau_{\rm eff}^\mu \to \infty$.  Noting that
$\lim_{\sigma \to 0} \Delta\left(\sigma,\rho \right) = 
2 \pi \delta\left(\rho\right)$, one finds that the stable orbit contribution
to $W_{osc}$ is proportional to the sum
$\sum_{\ell}\delta(\varepsilon - \varepsilon_{n,\ell})$.
The energies $\varepsilon_{n,\ell}$
are determined from the semiclassical quantization condition
of the longitudinal energy $\varepsilon_\ell^+ \equiv 
\varepsilon - \hbar \omega_\perp^{\mu,+} (\ell + 1/2)$
along the periodic orbit 
$S_\mu\left(\varepsilon_{n,\ell}\right)  = 2 \pi \hbar 
\left(n + \frac{n_\mu}{4}\right)$. Due to the harmonic approximation
the quantization of the transverse oscillations around the PO simply 
yields equally spaced levels with spacing $\hbar \omega_\perp^{\mu,+}$, 
where the frequency $\omega_\perp^{\mu,+} \equiv \phi_\mu/T_\mu$, with 
$\phi_{\mu}$ the winding number  \cite{gutzwiller} and $T_{\mu}$ the period 
of the orbit.  
We find that the coefficient functions in Eq. 
(\ref{w_po}) are the Wigner transforms of the harmonic oscillator
wavefunctions corresponding to these transverse modes:
\begin{eqnarray}
g_\ell^{\mu,+}\left(y, p_y\right) & \equiv & 
(-1)^\ell 
L_\ell\left(2 \left|Q_{\mu}\right|\right)
\exp\left(- \left|Q_{\mu}\right| \right)
\label{w_stable}
\end{eqnarray}
where $L_\ell$ is Laguerre polynomial and $Q_{\mu}$ is 
defined in (\ref{w_osc}).
Since the result (\ref{w_stable}) is
based on the harmonic approximation within a stable island, it is only
valid for quantum numbers $\ell$ such that the oscillations remain within
the island.  Equivalently, the number of such modes $\ell_{\rm max}$ 
is given by the ratio of the island area to $\hbar$.
The effect of phonon scattering according to (\ref{w_po}) 
is to smear out each of these 
delta-function contributions to $W_{osc}$ over a scale $\hbar/\tau_{opt}$.

(ii)  {\it Unstable orbits}.   For unstable periodic orbits 
we find that the ``effective'' relaxation rate
$1/\tau_{\rm eff}^\mu = 1/\tau_{\rm opt} + \left( \ell + 1/2 \right) 
\lambda_\mu$, where $\lambda_\mu$
is the Lyapunov exponent near the orbit $\mu$. Hence $\tau_{\rm eff}$
 is finite
and equal to the Lyapunov time when $\tau_{\rm opt} \to \infty$.
Therefore, instability acts as a sort of intrinsic level-broadening and
the contribution of unstable POs to (\ref{w_po}) never approaches a
delta function describing individual levels.
Instead this contribution describes the well-known clustering of levels 
responsible for Gutzwiller oscillations of the density of states
\cite{gutzwiller}, and in the tunneling rate each PO contribution is
weighted by the function
\begin{eqnarray}
g_\ell^{\mu,-}\left(y, p_y\right) & = &
(-1)^\ell
\Re\left\{
L_\ell\left(2 i Q_{\mu}\right)
\exp\left( i Q_{\mu} \right)
\right.
\nonumber \\
& \times & \left.
\left(1
+ i
\frac
{\sin\left(S_\mu/\hbar - \pi n_\mu/2 \right)}
{\sinh\left( T_\mu / \tau_{\rm eff}^\mu \right)}  
\right) \right\}
\label{w_unstable}
\end{eqnarray}
corresponding to Wigner functions
{\it averaged} \cite{berry} over the eigenstates of the cluster.

It is not clear in this case how to introduce the cut-off
value $\ell_{\rm max}$, however for each PO the contributions 
to $W_{\rm osc}$ decay exponentially 
$\sim \exp\left( - \lambda_\mu T_\mu \ell \right)$ with increasing 
$l$, and we find that for
our case the main contribution to the tunneling rate is given by the 
$\ell=0$ term.

Above we have argued that among the POs which reach the
emitter only those with points of contact within a distance 
$\sim l_B$ from the
injection point contribute strongly.  Our formulas (4),(5),(6) 
permit us now to propose a precise criterion : a PO is ``accessible''
to tunneling electrons if the emitter Wigner function centered on 
$y = y_i$, $p_y=0$ and of spatial width $\sim l_B$ overlaps
the functions $g^{\mu,\pm}$ centered at the contact points 
$y_{\mu},\left(p^{\mu}\right)_y$.
The localization length of the wavefunctions $\Psi_{\mu,\ell}^w$
corresponding to the functions $g^{\mu,+}$ can be shown by an extension of
our above analysis to be 
$ l_\mu  =  \sqrt{2 \hbar \left| m_{12}^{\mu}\right| }
/ \sqrt[4]{ \left| 4 - Tr^2\left[M_\mu\right] \right| } 
$.
The spatial scale associated with an unstable orbit is of the same form
but instead of Gaussian decay of the wavefunctions at 
$\delta y > l_\mu$ one finds rapid oscillations
 $\Psi_{\mu,\ell}^w \sim \exp\left( i \delta y^2/2 l_\mu^2\right)$.  

The relations (\ref{w_po}),(\ref{w_stable}),(\ref{w_unstable})
constitute a precise semiclassical expression for 
the tunneling rate in terms of the 
contributions of the distinct periodic orbits which reach the emitter
barrier (``emitter orbits'').   We will apply them now to a specific
parameter regime in the tilted well to compare qualitatively and
quantitatively to the experimentally-observed I-V characteristics.

In the recent periodic orbit theory \cite{ns_long} is was shown that
within the set of periodic orbits (POs) which 
collide with the collector $n$ times (period-$n$ orbits), there exist
orbits which collide with the emitter $m$ times, where $m \leq n$,
and it is useful to classify POs by the two integers $(m,n)$.
At $\theta =0$ the 
only resonances observed in the I-V characteristic are associated with
Bohr-Sommerfeld quantization of the $(1,1)$ orbit which traverses the
well with zero cyclotron energy.  When $\theta \neq 0$ such period-one
``traversing'' orbits still exist in some regions of the $B-V$
parameter space and define a background frequency of resonance 
peaks. However now additional resonances appear corresponding to
doubling or tripling of the frequency of peaks \cite{muller}.
These new peaks are associated with the 
existence of period-two and period-three orbits 
which appear and disappear as a result of bifurcations\cite{ns_long}.  
Here we focus on the peak-doubling in the interval $29^\circ
< \theta < 34^\circ$, where, as was shown in Ref. 5, 
there are four most relevant orbits, denoted by
$(1,2)_{1}$, $(1,2)_{2}$, $(1,2)^*_{1}$, $(1,2)^*_{2}$.  For
$\theta < 30^{\circ}$, the starred and unstarred orbits are paired
in the sense that they eventually become identical and disappear in
inverse tangent bifurcations.

The evolution of these orbits is represented by the four colored 
lines in Fig. 1.
We recall \cite{muller,ss,monteiro,ns_long} that under experimental
conditions the classical mechanics depends only on two
parameters $\beta = 2 B v_0 / E$ (where
$v_0$ is the velocity corresponding to the 
total energy : $v_0 \equiv \sqrt{2 \varepsilon/m^*}$) 
and on the tilt angle $\theta$.
As $\beta$ increases from zero, these four $(1,2)$ orbits
appear in cusp bifurcations \cite{ns_long,ns_scars,explaincusp}
and then disappear pairwise at higher $\beta$ in the inverse 
tangent bifurcations already mentioned.
The hatched region denotes the semiclassical width of the emitter
state, while the gray-scale regions denote the semiclassical widths
defined by the $g^{\mu}$ (evaluated here at $B = 8$ T) for the most 
accessible of these orbits for a given value of $\theta$ \cite{bifurcations}.  
Wherever these regions overlap 
the semiclassical formula (4) will predict
peak-doubling regions to appear in the $B-V$ parameter space
for the corresponding values of $\beta$, as seen in Fig. 2.

A fascinating feature of the classical dynamics, noted in Ref. 5, 
occurs near $\theta=30^{\circ}$ for the
experimental parameters of Ref. 2. 
The four $(1,2)$ orbits undergo an exchange bifurcation \cite{lichtenberg} 
so that for $\theta > 30^{\circ}$ the $(1,2)_1$ is paired with
$(1,2)_2^*$, whereas the $(1,2)_1^*$ orbit is now paired with the  
$(1,2)_2$.  For $\theta \approx 29^{\circ}$ the
$(1,2)_2$ orbit is accessible for over its entire interval of 
existence, $4.3 < \beta < 8.$ and the $(1,2)_1^*$
is accessible in the overlapping interval $7.9 < \beta <10.9$.
This situation manifests itself in a large and continuous 
region of peak-doubling in the experimental data\cite{muller}.
By $\theta = 31^\circ$, the pairing of the $(1,2)$ orbits has been
interchanged (Fig. 1b); the $(1,2)_1^*-(1,2)_2$ pair exists at lower
$\beta$ (higher voltage), and is moving away from the 
semiclassical accessibility region, whereas the $(1,2)_1$ orbit
has become most accessible.  Thus we expect the 
peak-doubling region to split into two smaller regions (Fig. 2b) and the
amplitude of the oscillations to be weaker
in the low-magnetic-field, high-voltage region (Fig. 2a). 
By $\theta=34^{\circ}$, the $(1,2)_2$ orbit 
has become inaccessible and these oscillations are no longer seen in Figs. 2d
and 2e. In contrast, the $(1,2)_1$ orbit
remains accessible and still produces strong
oscillations in the data (Figs. 2d,e).

In Figs. 2c and 2f, we show the results 
of the semiclassical theory
at $\theta = 31^\circ$ and $\theta = 34^\circ$.  
The semiclassical calculation of the tunneling current
using formulae (\ref{w_po}),(\ref{w_stable}),
(\ref{w_unstable}) is in good agreement with the experimental data
\cite{endnote}.

We acknowledge helpful conversations with Gregor Hackenbroich, 
Tania Monteiro and John Delos. The work of A.D.S. and E.N.
was supported by NSF grant DMR-9215065.

\onecolumn

\begin{figure}[htbp]
\begin{center}
\leavevmode
\epsfxsize = 6in
\epsfbox{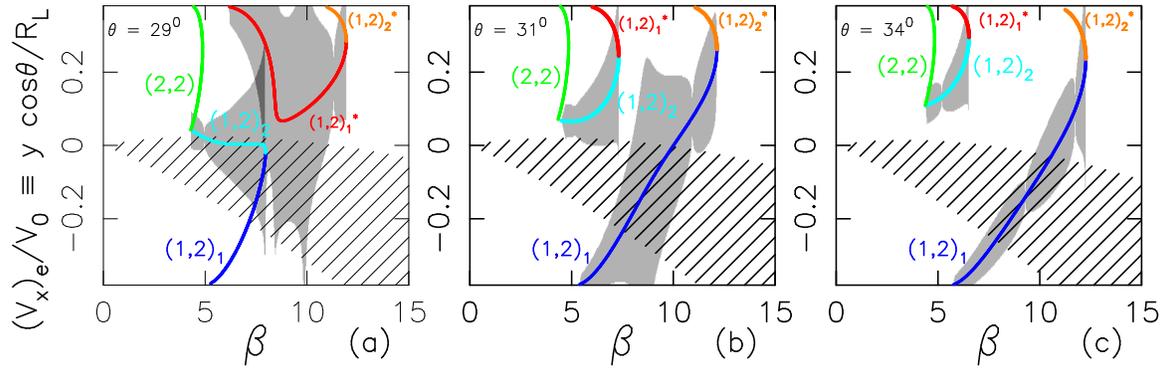}
\end{center}
\caption{ (color)
Bifurcation diagrams for the relevant period-two orbits at (a) before 
($\theta = 29^\circ$) and (b) after ($\theta = 31^\circ$ and $34^\circ$) 
the 
exchange bifurcation.
The shading represents the calculated localization lengths of 
the wavefunctions localized near the most relevant $(1,2)$ periodic orbits 
(see text). 
The hatched region denotes the semiclassical width of the emitter
state.
\label{bif_diag} 
}
\end{figure}

\begin{figure}[htbp]
\begin{center}
\leavevmode
\epsfxsize = 6 in
\epsfbox{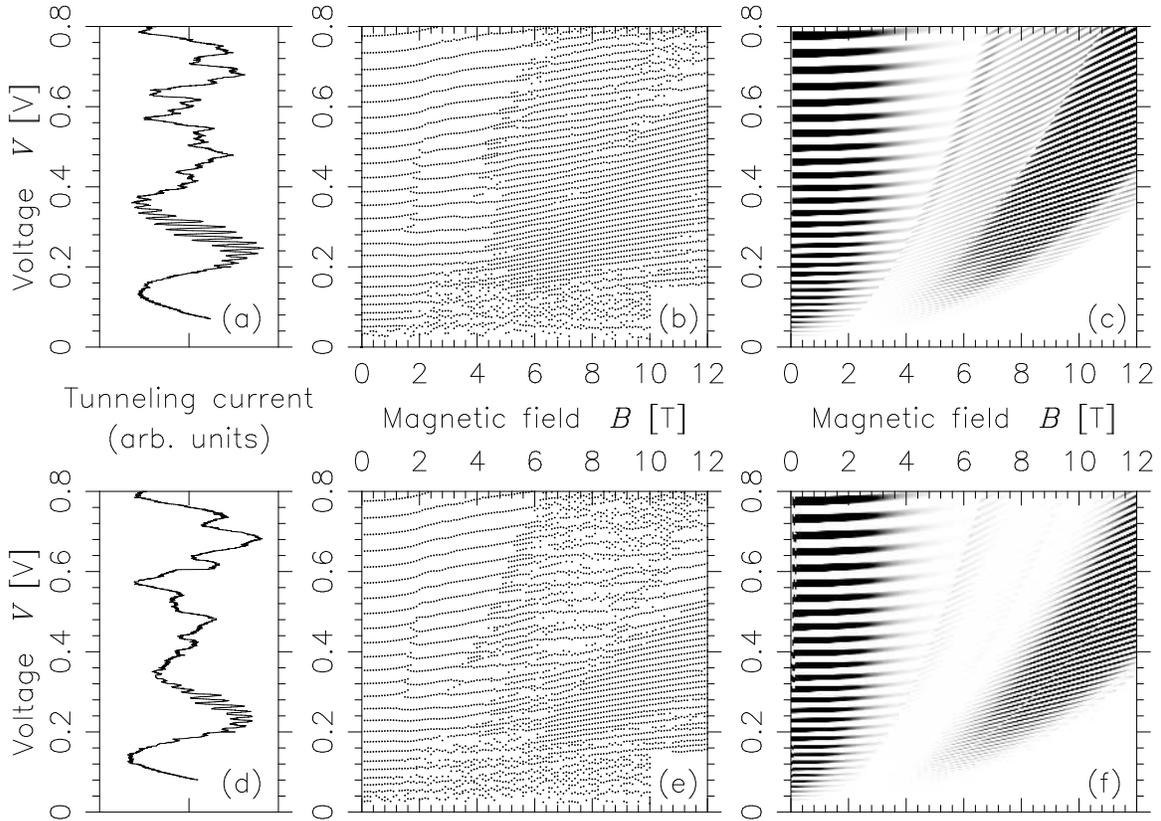}
\end{center}
\caption{
(a),(d) Samples of experimental resonant tunneling $I-V$ traces, from which 
peak
positions are determined and (b),(e) plotted versus total magnetic field, 
to be compared with (c),(f) results of the semiclassical calculation.
Note at (a-c) $\theta = 31^\circ$ 
the onset of the separation of the large region of peak-doubling
immediately after the exchange bifurcation and, at (d-f) $\theta = 
34^\circ$, the peak-doubled region at lower magnetic field created by the exchange bifurcation
is almost invisible due to much weaker coupling of the corresponding 
orbit $(1,2)_2$ to the emitter (see Fig. 1c). The I-V traces (a), (d) 
evidence this by the disappearance of the oscillations at higher bias
voltages.
\label{data} 
}
\end{figure}


\begin{references}
\bibitem{explain_emitter_occupation} The equation in the text corresponds
to the case when only one of the 2DEG energy levels is occupied. 
The generalization to several occupied levels is straightforward.
\bibitem{explaincusp} A cusp bifurcation is a new kind of bifurcation
which occurs in this system due to the nonanalyticity of the Poincar\'e
map at the emitter barrier, see Ref. 5.
\bibitem{bifurcations}  The semiclassical formula of Eq. (4) breaks
down at bifurcations, when the periodic orbits are not isolated.
Correspondingly, the semiclassical width can either diverge or
vanish (depending on the nature of the bifurcation).  These divergences
are spurious but are integrable in Eq. (4) 
and only lead to inaccuracies in a small
interval of $\beta$ very close to the bifurcation.
\bibitem{endnote}
The major remaining discrepancy, that the experimental 
peak-doubling regions are observed at lower voltages than calculated
by the semiclassical theory, arises from an offset voltage in the 
sample\protect\cite{muller} and band non-parabolicity effects 
\protect\cite{scars}.
\end{references}
\end{document}